\documentclass[10pt,journal,compsoc]{IEEEtran}

\usepackage{graphicx}
\usepackage{subfigure}
\usepackage{multirow}

\usepackage{amsmath}   
\allowdisplaybreaks[4]   

\ifCLASSOPTIONcompsoc
  \usepackage[nocompress]{cite}
\else
  \usepackage{cite}
\fi

\usepackage{xcolor}
\usepackage{multirow}

\begin{document}

\title{Discovery of cancer common and specific driver gene sets}

\author{Junhua~Zhang and~Shihua~Zhang 
\IEEEcompsocitemizethanks{\IEEEcompsocthanksitem J. Zhang and S. Zhang are with National Center for Mathematics and Interdisciplinary Sciences, Academy of Mathematics and Systems Science, Chinese Academy of Sciences, Beijing 100190, China. \protect\\
E-mail: zjh@amt.ac.cn, zsh@amss.ac.cn.}
}


%


\IEEEtitleabstractindextext{%
\begin{abstract}
\noindent Cancer is known as a disease mainly caused by gene alterations. Discovery of mutated driver pathways or gene sets is becoming an important step to understand molecular mechanisms of carcinogenesis. However, systematically investigating commonalities and specificities of driver gene sets among multiple cancer types is still a great challenge, but this investigation will undoubtedly benefit deciphering cancers and will be helpful for personalized therapy and precision medicine in cancer treatment. In this study, we propose two optimization models to \emph{de novo} discover common driver gene sets among multiple cancer types (ComMDP) and specific driver gene sets of one certain or multiple cancer types to other cancers (SpeMDP), respectively. We first apply ComMDP and SpeMDP to simulated data to validate their efficiency. Then, we further apply these methods to 12 cancer types from The Cancer Genome Atlas (TCGA) and obtain several biologically meaningful driver pathways. As examples, we construct a common cancer pathway model for BRCA and OV, infer a complex driver pathway model for BRCA carcinogenesis based on common driver gene sets of BRCA with eight cancer types, and investigate specific driver pathways of the liquid cancer lymphoblastic acute myeloid leukemia (LAML) versus other solid cancer types. In these processes more candidate cancer genes are also found.
\end{abstract}

}

\maketitle

\IEEEdisplaynontitleabstractindextext

%
\IEEEpeerreviewmaketitle


\section{Introduction}
\IEEEPARstart{C}{ancer} is a complex and heterogeneous disease with diverse genetic and environmental factors involved in its etiology. With the advances of deep sequencing technology, huge volume cancer genomics data have been generated through several large-scale programs (e.g., The Cancer Genome Atlas (TCGA) \cite{TCGA}, International Cancer Genome Consortium (ICGC) \cite{ICGC}, and the Cancer Cell Line Encyclopedia (CCLE) \cite{Barretina}), which provide huge opportunities for understanding the molecular mechanisms and pathogenesis underlying cancer \cite{Zhang2012}. Currently, a crucial challenge in cancer genomics is to distinguish driver mutations and driver genes which contribute to cancer initiation and development from passenger ones which accumulate in cells but do not contribute to carcinogenesis \cite{Greenman,Stratton}. Most early efforts have been devoted to detect individual driver genes with recurrent mutations \cite{Beroukhim}. However, this kind of methods do not consider the complicated mutational heterogeneity in cancer genomes with diverse mutations in genes.

Although cancer patients exhibit diverse genomic alterations, many studies have demonstrated that driver mutations tend to affect a limited number of cellular signaling and regulatory pathways \cite{Vogelstein_04,TCGA,Ding}. Therefore, a great deal of attention has been devoted to evaluate the recurrence of mutations in groups of genes derived from known pathways or protein-protein interaction networks \cite{Ding,Jones,Ciriello}. These groups of genes are considered as candidate driver pathways, which may be frequently perturbed within tumor cells \cite{Boca,Efroni} and can lead to the acquisition of carcinogenic properties such as cell proliferation, angiogenesis, or metastasis \cite{Hanahan_00,Hanahan_11}. A main concern is that the human protein interaction network and biological pathways are far from being complete. It is necessary to develop new methods without relying on prior knowledge to discover novel mutated driver gene sets or pathways.


Previous studies indicate that a driver gene set has two key properties: (1) covering a large number of samples (high coverage); and (2) its mutations tend to exhibit mutual exclusivity (high mutual exclusivity), i.e., a single mutation is usually enough to disturb one pathway \cite{Vogelstein_04, Yeang,Zhang_2016}. For example, the mutation of \emph{TP53} and the copy number amplification of \emph{MDM2} seldom appear simultaneously in glioblastoma multiforme (GBM) patients (p53 pathway) \cite{TCGA}. These rules have been frequently used to \emph{de novo} discover mutated driver gene sets in recent years \cite{Vandin,Zhao,Zhang}. For example, Vandin \emph{et al.} developed Dendrix  by designing a weight function to combine the coverage and exclusivity of a gene set, and maximizing it via a Markov chain Monte Carlo (MCMC) approach to extract driver gene sets \cite{Vandin}. Zhao \emph{et al.} further developed a binary linear programming (BLP) model \cite{Zhao} to get the exact solutions of the maximization problem, and designed a genetic algorithm to optimize variant weight functions and incorporate prior biological knowledge into it in a more flexible manner. However, these studies have all focused on a single pathway without considering the cooperativeness between pathways \cite{Vandin,Zhao,Zhang,Miller}.

In fact, a great deal of evidence has suggested that pathways often function cooperatively in cancer initiation and progression \cite{Hanahan_11,Yeang,Cui,Klijn}. Thus, exploring the complex collaboration among different biological pathways and functional modules may shed new lights on the understanding of the cellular mechanisms underlying carcinogenesis. Leiserson \emph{et al.} \cite{Leiserson_13} generalized Dendrix (Multi-Dendrix) to simultaneously identify multiple driver gene sets in cancer. More importantly, the collaboration among different pathways means these gene sets are likely simultaneously mutated in a large cohort of patients. To this end, Zhang \emph{et al.} \cite{Zhang_2014} developed CoMDP to \emph{de novo} discover co-occurring mutated driver gene sets in cancer by introducing a novel weight function and a mathematical programming model; Melamed \emph{et al.} \cite{Melamed} introduced an information theoretic method GAMToC to identify combinations of genomic alterations in cancer; and Remy \emph{et al.} \cite{Remy_2015} developed a logical model to explain mutually exclusive and co-occurring genetic alterations in bladder carcinogenesis.

On the other hand, different cancer types may have certain commonalities \cite{Kandoth}. Investigating the similarities and differences among multiple cancer types may enhance the understanding of pathologies underlying cancers and provide new clues to efficient drug design and cancer treatment. The TCGA pan-cancer project surveyed multi-platform aberration data in cancer samples from thousands of cancer patients among 12 cancer types \cite{Weinstein_13}, which provides huge opportunities to make such investigations \cite{Ciriello_13,Liu2014}. For example, different histological cancers can be classified into the same clusters \cite{Ciriello_13,Hofree,Liu2015}, which means that different cancers may be treated by the same drugs. Recently, Leiserson \emph{et al.} \cite{Leiserson_15} proposed a directed heat diffusion model (HotNet2) to identify pathways and protein complexes based on pan-cancer network analysis; Kim \emph{et al.} \cite{Kim_15} investigated different kinds of mutual exclusivity among multiple cancer types and designed statistical testing methods for driver gene set identification (MEMCover). Although recent pan-cancer studies revealed that some pairs of genes showing mutually exclusivity are common or specific for some cancer types \cite{Kandoth,Szczurek}, there is still a lack of systematic investigation of commonalities and specificity in pathway level.

In this study, we develop two mathematical programming models (ComMDP and SpeMDP) to \emph{de novo} identify cancer common and specific driver gene sets, respectively. For the former, we detect a set of genes which have significantly high mutual exclusivity and large coverage in two or more cancer types simultaneously. For the latter, we identify a driver gene set specific to one or a group of cancer types (say, $S_1$) versus another group of cancer types (say, $S_2$). In other words, we require the detected genes to have significantly high mutual exclusivity and large coverage in the group $S_1$ but not in $S_2$. We first apply ComMDP and SpeMDP to simulated data to validate their effectiveness. Then, we apply them to the mutation data of 12 cancer types from the pan-cancer project \cite{Weinstein_13,Ciriello_13} and obtain several biologically meaningful driver gene sets. For example, for breast carcinoma (BRCA) and ovarian carcinoma (OV), we identify their common driver gene sets as well as their individual specific driver gene sets relative to the other. Interestingly, the identified common gene sets are involved with distinct cancer pathways such as apoptosis pathway, \emph{ErbB} signaling pathway, \emph{PI3K-Akt} signaling pathway and \emph{MAPK} signaling pathway, which enable us to construct a common cancer pathway model for BRCA and OV. Further, we construct a hypothetical mutated driver pathway model for BRCA carcinogenesis and progression based on eight common driver gene sets of BRCA with eight cancer types, indicating the complexity of BRCA carcinogenesis. In addition, we investigate specific driver pathways of the liquid cancer lymphoblastic acute myeloid leukemia (LAML) versus other solid cancer types, and identify mutations of \emph{FLT3, IDH2, NRAS, IDH1, RUNX1, NPM1, TET2, KIT}, amplifications of \emph{MLL, IGSF5}, and deletions of \emph{TP53, GNAQ}, which are involved in proliferation, transcriptional deregulation, impaired hematopoietic differentiation, and so on. We expect the proposed methods can discover new commonalities and specificities among cancers and help to understand cancer initialization and progression further.

\section{MATERIALS AND METHODS}

\begin{figure}[!t]
\centering
\includegraphics[width=3.4in]{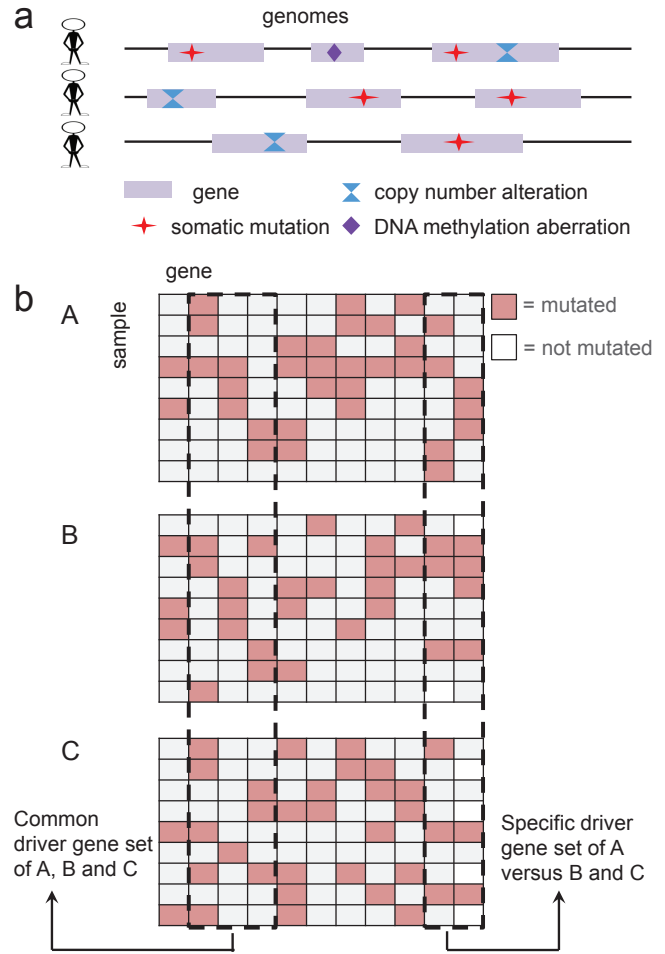}
\caption{Schematic illustration of the key idea of this study. (a) Obtain the mutation matrix from the sequencing data (referring to somatic mutations, copy number alterations (CNAs) and DNA methylation aberrations) \cite{Vandin}. (b) Identify the common and specific driver gene sets using ComMDP and SpeMDP.}
\label{fig1}
\end{figure}

We first briefly describe the maximum weight submatrix problem, where the coverage and exclusivity of a gene set is combined to form a weight function for discovering driver gene sets in a single mutation data \cite{Vandin,Zhao}. Then we propose ComMDP and SpeMDP to \emph{de novo} discover cancer common and specific mutated driver gene sets among multiple cancer types, respectively (Fig. \ref{fig1}).

\subsection{The maximum weight submatrix problem}
Given a binary mutation matrix $A$ with $m$ rows (samples) and $n$ columns (genes), Vandin \emph{et al.} introduced a weight function $W$ and defined the maximum weight submatrix problem \cite{Vandin}. Specifically, it is designed to find a submatrix $M$ of size $m\times k$ in matrix $A$ by maximizing $W$:
\begin{equation}\label{eq:1}
W(M)=|\Gamma (M)|-\omega (M)=2|\Gamma (M)|-\sum_{g\in M}|\Gamma (g)|,
\end{equation}
where $\Gamma (g)=\{i: A_{ig}=1\}$ denotes the set of samples in which the gene $g$ is mutated, $\Gamma (M)=\cup_{g\in M} \Gamma (g)$ measures the coverage of $M$, and $\omega (M)=\sum_{g\in M}|\Gamma (g)|-|\Gamma (M)|$ measures the coverage overlap of $M$.

\subsection{ComMDP for identifying common mutated driver gene sets among two or multiple cancer types}

Considering $R$ ($R\geq 2$) cancer types, for each we have the mutation matrix $A_r=(a_{ij}^{(r)})$ with $m_r$ samples and the same $n$ mutated genes $(r=1, \ldots, R)$. To find a common mutated driver gene set $M$ with large coverage and high exclusivity, we introduce a weight function $C_m$:
\begin{equation}\label{eq:2}
C_m(M)=\displaystyle \sum_{r=1}^{R} \big [2|\Gamma_{A_r} (M)|-\sum_{g\in M}|\Gamma_{A_r} (g)| \big ].
\end{equation}
We propose the following BLP model to maximize it:
\begin{equation}\label{eq:3}
\begin{split}
\max \;\; F_m(\textbf{x},u)= & \displaystyle  \sum_{r=1}^R \Big [ 2\sum_{i=1}^{m_r}x_i^{(r)}-\sum_{j=1}^{n}
                \Big ( u_j\cdot \sum_{i=1}^{m_r}a_{ij}^{(r)}\Big ) \Big]   \\
\end{split}
\end{equation}
\begin{equation}\label{eq:4}
\mbox{s.t.} \;\; \left\{ \;\;
\begin{array}{l}
\vspace{1mm} \displaystyle
x_i^{(r)}\leq \sum_{j=1}^n a_{ij}^{(r)}u_j, \; i=1, \ldots, m_r, \; r=1, \ldots, R, \\
\vspace{1mm}
\displaystyle \sum_{j=1}^n u_j=K, \\
\begin{split}
\displaystyle x_i^{(r)}, u_j \in \{0, 1\}, \; & i=1, \ldots, m_r, \; j=1, \ldots, n, \\
& r=1, \ldots, R,
\end{split}
\end{array}
\right.
\end{equation}
where $u_j$ indicates whether column $j$ of the mutation matrices falls into submatrx $M$ or not, and all the columns $j$'s with $u_j=1$ constitute $M$; $\textbf{x}=\{x^{(1)},\ldots,x^{(R)}\}$, and $x_i^{(r)}$ indicates whether the entries of row $i$ are zeros or not in $A_r$ ($r=1, \ldots, R$). Thus, $\sum_{i=1}^{m_r} x_i^{(r)}$ represents the coverage of $M$ in $A_r$ (i.e., $|\Gamma_{A_r} (M)|$); $K$ is the total number of genes within $M$.

\subsection{SpeMDP for identifying a certain or multiple cancer specific driver gene sets}

Suppose we want to find the specific mutated driver gene sets for $R$ cancer types relative to other $T$ ones ($R\geq 1, T\geq 1$). We use $A_r=(a_{ij}^{(r)})$ $(r=1, \ldots, R)$ and $B_t=(b_{kj}^{(t)})$ $(t=1, \ldots, T)$ to denote corresponding mutation matrices, respectively. We introduce the weight function $S_{m}$:
\begin{equation}\label{eq:5}
\begin{split}
S_{m}(M)= & \displaystyle \frac{1}{R}\sum_{r=1}^{R} \big [K|\Gamma_{A_r} (M)|-\sum_{g\in M}|\Gamma_{A_r} (g)| \big ] \\
& -\displaystyle \frac{1}{T}\sum_{t=1}^{T} \big [K|\Gamma_{B_t} (M)|-\sum_{g\in M}|\Gamma_{B_t} (g)| \big ].
\end{split}
\end{equation}
We maximize $S_{m}$ by the following BLP model:
\begin{equation}\label{eq:6}
\begin{split}
\max \;\; & G_{m}(\textbf{x},\textbf{y},u)  =\displaystyle \frac{1}{R}\sum_{r=1}^{R} \Big [K\sum_{i=1}^{m_r} x_i^{(r)}-\sum_{j=1}^{n} \Big (u_j\cdot \sum_{i=1}^{m_r} a_{ij}^{(r)}\Big ) \Big ] \\
& \;\;\;\; - \frac{1}{T}\sum_{t=1}^{T} \Big [K\sum_{k=1}^{l_t}y_k^{(t)}-\sum_{j=1}^{n}
                \Big (u_j\cdot \sum_{k=1}^{l_t}b_{kj}^{(t)}\Big ) \Big ],
\end{split}
\end{equation}
\begin{equation}\label{eq:7}
\mbox{s.t.} \;\; \left\{ \;\;
\begin{array}{l}
\vspace{1mm} \displaystyle
x_i^{(r)}\leq \sum_{j=1}^n a_{ij}^{(r)}u_j, \;\;\; i=1, \ldots, m_r, \;\; r=1, \ldots, R, \\
\vspace{1mm}
\begin{split}
\displaystyle
\frac{1}{n}\sum_{j=1}^n b_{kj}^{(t)}u_j & \leq y_k^{(t)}\leq \sum_{j=1}^n b_{kj}^{(t)}u_j, \\
& k=1, \ldots, l_t, \;\; t=1, \ldots, T,
\end{split}
\\
\displaystyle \sum_{j=1}^n u_j=K, \\
\displaystyle x_i^{(r)}, y_k^{(t)}, u_j \in \{0, 1\}, \; i=1, \ldots, m_r, \;\; j=1, \ldots, n, \\
\quad\quad\quad\quad r=1, \ldots, R, \; k=1, \ldots, l_t, \; t=1, \ldots, T,
\end{array}
\right.
\end{equation}
where $\textbf{x}=\{x^{(1)},\ldots,x^{(R)}\}, \textbf{y}=\{y^{(1)},\ldots,y^{(T)}\}$. As stated above, the constraint $x_i^{(r)}\leq \sum_{j=1}^n a_{ij}^{(r)}u_j$ in Eq. (\ref{eq:7}) ensures that $\sum_{i=1}^{m_r} x_i^{(r)}$ is the coverage of $M$ in $A_r$. In Eq. (\ref{eq:5}) or Eq. (\ref{eq:6}), because of the subtraction of the weights of $B_t$ from that of $A_r$, we use the restrictions $\frac{1}{n}\sum_{j=1}^n b_{kj}^{(t)}u_j\leq y_k^{(t)}\leq \sum_{j=1}^n b_{kj}^{(t)}u_j$ to ensure that $\sum_{k=1}^{l_t} y_k^{(t)}$ is the coverage of $M$ in $B_t$, and we use the coefficient $K$ to ensure the weights of $A_r$ and $B_t$ are all non-negative.

\subsection{Statistical significance}

We perform a permutation test to assess the significance of results. We permutate the mutations independently among samples to preserve the mutation frequency of each gene. Two kinds of significance are calculated: (1) individual one measuring the significance of a gene set in a certain mutation matrix, where the weight $W$ in Eq. (\ref{eq:1}) is used as the statistic; (2) overall one measuring the significance of a gene set by viewing all the mutation matrices as a whole, where the weight $C_m$ in Eq. (\ref{eq:2}) and the weight $S_m$ in Eq. (\ref{eq:5}) are used as the statistics for ComMDP and SpeMDP, respectively.

\subsection{Simulated data}

To assess the performance of the proposed methods on a variety of data, we construct eight datasets, sd1, $\cdots$, sd8, for simulation study. For convenience of description, in the following we use $A_r$ or $B_r$ to denote the mutation matrices, $M_i^{(r)}$ to denote the $i$-th embedded submatrix (or gene set) in $A_r$ or $B_r$ for which the proposed methods are used to identify, and $p_i^{(r)}$ to denote the gene mutation rate in $M_i^{(r)}$ ($1\leq r\leq R, 1\leq i\leq I$).

The datasets sd1 and sd2 are generated to illustrate the performance of ComMDP for identifying common driver gene sets among multiple cancer types. The difference is that in sd1 for each $r$ the $M_i^{(r)}$ have a constant mutation rate ($1\leq i\leq I$), but in sd2 they have varying ones, to investigate the possible impact of mutation rates in the gene sets on the discovery accuracy. sd1 is constructed as follows. First, we have three empty matrices $A_r$ with the same sizes: $m$ (samples) $\times$ $n$ (genes) (here $m=500$, $n=900$). Then, we embed $I$ submatrices $M_i^{(r)}$ with a mutation rate $p^{(r)}$ into each matrix $A_r$ ($r=1,\cdots,3$; $i=1,\cdots,I$; $I=9$; $p^{(1)}=0.80$, $p^{(2)}=0.85$, $p^{(3)}=0.90$), where for each $r$, $M_i^{(r)}$ contains $i+1$ genes ($i=1,\cdots,I$), and for each $i$, these submatrices $M_i^{(r)}$ occupy the same columns in the corresponding $A_r$ $(r=1,\cdots,3)$. For each sample in $A_r$, a gene uniformly chosen from $M_i^{(r)}$ is mutated with rate $p^{(r)}$, and once one gene is mutated, the other genes in $M_i^{(r)}$ have a rate $p_0$ to be mutated ($p_0=0.04$). Finally, the genes not in $M_i^{(r)}$ are mutated in at most three samples, which can be viewed as the background mutation rate in the simulated data. The dataset sd2 is constructed in a similar way, the difference is that each gene set has 9 genes ($K=9$), and $M_i^{(r)}$ has a mutation rate $p_i^{(r)}=1-i*\delta (r) (r=1,\cdots,3; i=1,\cdots,9)$, where $\delta (1)=0.03, \delta (2)=0.04, \delta (3)=0.05$.

The simulated datasets sd3-sd7 are generated to demonstrate the performance of SpeMDP for identifying specific driver gene sets of one or several cancer type(s) versus other cancers. In this case, the datasets are constructed to contain two kinds of embedded gene sets. The first kind of gene sets have mutations with (approximately) mutual exclusivity, like those in sd1 (called the first manner of embedding); but for the second ones, we randomly select 60\% samples for which two genes are randomly chosen to be mutated with proper mutation rates, ensuring they are not exclusive (called the second manner of embedding). For the details of the construction of sd3-sd7, please refer to the Supplementary Data. We use these five datasets sd3-sd7 to investigate different aspects for SpeMDP applications. Specially,

- sd3 for discovering a certain cancer specific driver gene sets

- sd4 for investigating in which case SpeMDP can identify specific driver gene sets

- sd5 for investigating the impact of diverse mutation rates on the results

- sd6 for investigating the method performance in different mutation generation manners

- sd7 for discovering multiple cancer specific driver gene sets

\vskip1.6mm

We construct dataset sd8 to see if the previous individual cancer type approaches can also identify cancer common and specific driver gene sets (e.g., BLP in MDPFinder \cite{Zhao} or Dendrix \cite{Vandin}). The construction of sd8 is similar to that of sd1. Seven groups of submatrices $M_i^{(r)}$ with 6 genes in each are embedded into $A_r$ with $m=500, n=700, r=1,\cdots,3$. The first group of submatrices are constructed in the first embedding manner (thus corresponding to a common driver gene set); each of the 2nd to the 4th groups contains two with the first embedding manner, one with background mutations (corresponding to neither common nor specific driver gene sets); each of the last three groups contains one with the first embedding manner, two with background mutations (each corresponding to a specific driver gene set).

\subsection{Biological data}

We use mutation data from the pan-cancer project \cite{Ciriello_13,Liu2015} to assess our methods for practical applications. The 12 types of cancer include bladder carcinoma (BLCA), breast carcinoma (BRCA), colon adenocarcinoma (COAD), glioblastoma multiformae (GBM), head and neck squamous carcinoma (HNSC), kidney renal clear-cell carcinoma (KIRC), lymphoblastic acute myeloid leukemia (LAML), lung adenocarcinoma (LUAD), lung squamous carcinoma (LUSC), ovarian carcinoma (OV), rectal adenocarcinoma (READ), and uterine cervical and endometrial carcinoma (UCEC). Here colon adenocarcinoma and rectal adenocarcinoma are combined into one type denoted as COADREAD.

\section{RESULTS}

\subsection{Simulation study}

We first apply ComMDP to the simulated datasets sd1, sd2 and apply SpeMDP to sd3 - sd7 to assess their performance. We run each method ten times for each dataset. We further apply them to sd8 and compare them with driver gene set discovery approaches for individual cancer types (BLP \cite{Zhao} is used here).

\subsubsection{Common driver gene set discovery.}

\begin{figure*}[!t]
\centering
\includegraphics[width=5.7in]{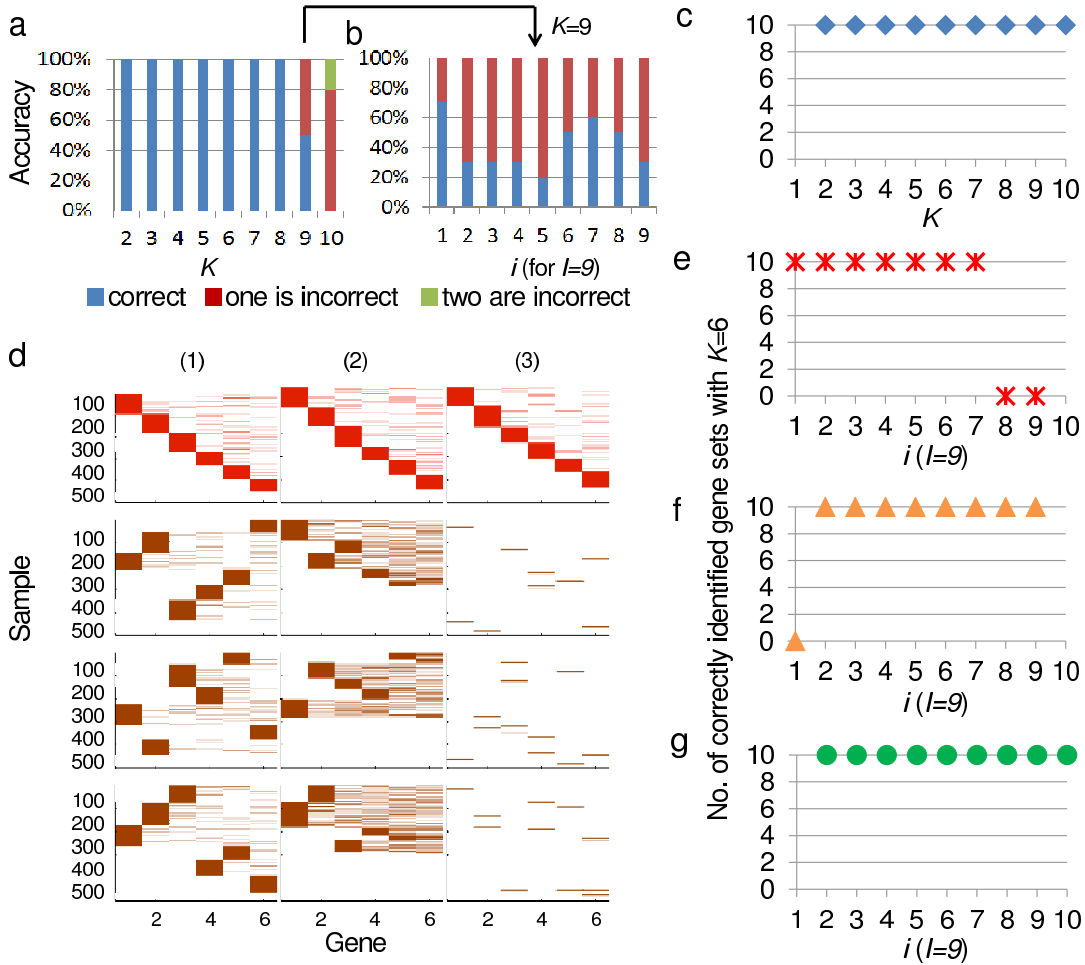}
\caption{The results of simulation study. Accuracy of the common driver gene set discovery: (a) for $K=$2 to 10 with a constant mutation rate and (b) for fixed $K=9$ in which the embedded nine gene sets ($I=9$) have decreasing mutation rates. (c) Accuracy of the specific driver gene set discovery for $K=$2 to 10 with a constant mutation rate. (d) Illustration of the situation for specific driver gene set discovery: (1) two kinds of exclusive gene sets corresponding to a common driver pathway, so it cannot be found; (2) exclusive (red) versus nonexclusive (brown) gene sets can be found; (3) exclusive (red) versus random (brown) gene sets can be found. Numbers of correctly identified specific driver gene sets for ten runs with $K=6$ and $I=9$: (e) both kinds of gene sets have decreasing mutation rates and (f) one has increasing and the other has decreasing mutation rates. (g) Numbers of correctly identified specific driver gene sets for ten runs about multiple cancer types with $K=6$ and $I=9$.}
\label{fig2}
\end{figure*}

For sd1 ComMDP can identify the embedded gene sets for all the ten runs when the number of genes $K\leq 8$ (Fig. \ref{fig2}a). When $K=9$, it can detect the embedded gene set for five runs, and it has a wrong one for each of other five runs. When $K=10$, each detection of eight runs contains nine correct genes plus a wrong one, and each of other two runs has two wrong ones. We also investigate the possible impact of varying mutation rates in the gene sets on the discovery accuracy. For the embedded nine gene sets of $K=9$ with diverse mutation rates in sd2, each of ten runs can identify at least eight correct genes in each gene set (Fig. \ref{fig2}b).

\subsubsection{Specific driver gene set discovery.}

First, we consider the situation of one cancer specific driver gene sets. In sd3, SpeMDP can correctly detect the embedded gene sets for $K=$2 to 10 (Fig. \ref{fig2}c). The results on sd4 demonstrate that when the gene set is exclusive in kind 1 (kind 2) set but not in kind 2 (kind 1) set, or one is exclusive and the other takes background mutations, SpeMDP can successfully identify it (Fig. \ref{fig2}d). In the first case, the gene set is approximately exclusive in both kinds 1 and 2 sets which corresponds to a common driver pathway, so SpeMDP cannot find it. In sd5, the mutation rates in kind 1 and kind 2 sets simultaneously get smaller and smaller, so the mutation coverage will get small (so does the weight $W$) in kind 1 set along with $i$ gets large ($1\leq i\leq I, I=9$), and more exclusive mutation in kind 2 dataset will become possible. Therefore, the performance to detect the embedded gene set will decrease when $i$ gets large (Fig. \ref{fig2}e). We further validate this on sd6. The mutation rate of sd6 in kind 2 dataset gets larger and that in kind 1 set gets smaller along with $i$ becomes large ($1\leq i\leq I, I=9$). In this case, we successfully identify all the embedded gene sets except the one corresponding to $i=2$ (Fig. \ref{fig2}f). Lastly, the result on sd7 indicates that SpeMDP is also effective to identify specific gene sets for multiple cancer types (Fig. \ref{fig2}g), where the dataset is simulated in a similar way to that of sd6.

\subsubsection{Individual driver gene set discovery approaches cannot detect common and specific driver gene sets well.}

\begin{figure*}[!t]
\centering
\includegraphics[width=5.5in]{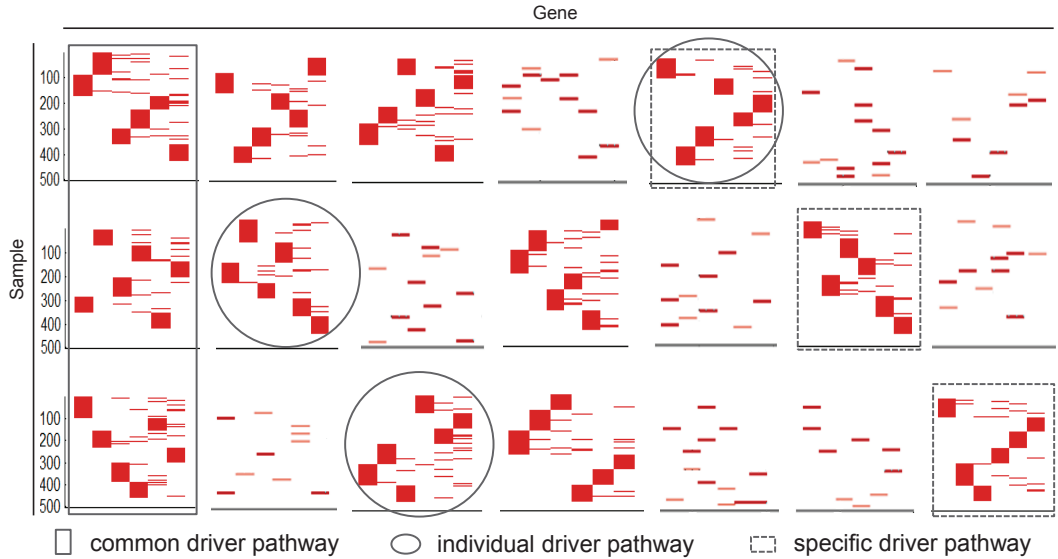}
\caption{ComMDP and SpeMDP can exactly identify the embedded common driver gene sets (rectangle) and specific driver gene sets (dotted rectangles), respectively. The BLP of MDPFinder \cite{Zhao} can identify the individual driver gene sets in each dataset (ellipses).}
\label{fig3}
\end{figure*}

For sd8, we first use the BLP model in MDPFinder \cite{Zhao} to identify individual driver gene sets in each $A_r$ which contains seven embedded submatrices, and we get the ones marked by ellipses in Fig. \ref{fig3}. Then we apply ComMDP and SpeMDP to identify the common and specific driver gene sets among all the $A_r$s, and we obtain those marked by the rectangle and dotted rectangles, respectively. Note that the detected individual and common driver gene sets do not have any overlap. Moreover, the detected individual driver pathways in the second and third sets ($A_2$ and $A_3$) are not specific (Fig. \ref{fig3}).

\subsection{Applications to biological data}

We investigate common driver gene sets among all the pair mutation data of the 11 cancer types with $K=$ 2 to 10. We summarize all the significant driver gene sets with both individual and overall significance less than 0.05 (Suppl. Table S1). The mutation rates of \emph{TP53} in LUSC and OV are very high (164/182 and 405/445, respectively). We distinguish this situations with or without \emph{TP53} when relating to these two cancer types.

\vskip5mm
\noindent{\it\textbf{Common mutated driver gene sets among two or multiple cancer types}}
\vskip2mm

\begin{figure*}[!t]
\centering
\includegraphics[width=5.1in]{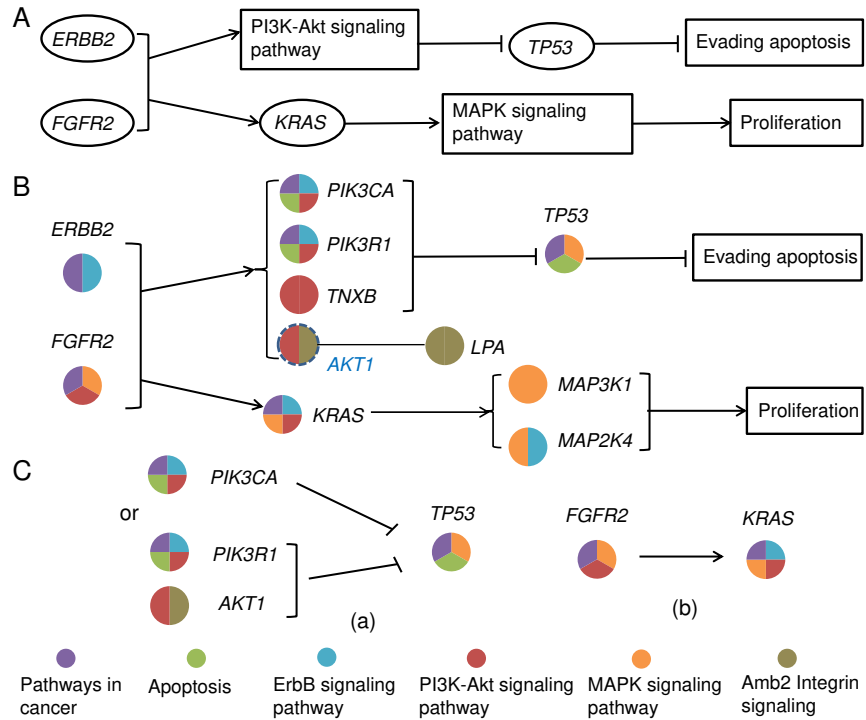}
\caption{(A) Known KEGG pathway knowledge in cancer. (B) A common mutated pathway model for BRCA and OV initiation and progression. It is inferred based on the identified common gene sets and their participant pathway knowledge. The gene \emph{AKT1} relating to LPA with the PI3K-Akt signaling pathway does not appear in the identified common gene sets. (C) Only local parts of the common mutated pathway in (B) can be found by the individual cancer type approach BLP: (a) BRCA (up for $2\leq K\leq 6$, below for $7\leq K\leq 10$), (b) OV.}
\label{fig4}
\end{figure*}

Previous studies indicate that BRCA and OV have similar phenotypes to some extent. Interestingly, we indeed obtain significant common driver gene sets between them by ComMDP for $K=$ 7 to 10 (Table \ref{table2}), and reveal 10 genes \emph{TP53, PIK3CA, MAP3K1, MAP2K4, PIK3R1, LPA, KRAS, ERBB2, FGFR2, TNXB} in total. These genes are enriched in several signaling pathways relating to apoptosis, ErbB signaling pathway, PI3K-Akt signaling pathway, MAPK signaling pathway, etc. Based on known KEGG pathway knowledge (Fig. \ref{fig4}A), we propose a common mutated pathway model for cancer initiation and progression in both BRCA and OV (Fig. \ref{fig4}B). We show the heat map of the alterations of the gene set for $K=10$ (Suppl. Figure S1) and see that \emph{TP53} has a very high mutation rate in OV (as stated above). The mutation rates of other nine genes are very low. It implies that \emph{TP53} mutation plays a dominant role in this pathway in OV, indicating that the common driver pathway exploration helps to identify driver ones with low mutation frequency (Fig. \ref{fig4}).
\begin{table*}[!ht]
\renewcommand{\arraystretch}{1.3}
\caption{Significant common driver gene sets between BRCA and OV identified by BLP}
\label{table2}
\small
\centering
      \begin{tabular}{|l|l|l|l|l|}
        \hline
        $K$ & Common pathways & $p_1$ & $p_2$ & $p$ \\ \hline

        7 & \emph{TP53, PIK3CA, MAP3K1, MAP2K4, PIK3R1, LPA, KRAS} & 0 & 0.003 & 0 \\ \hline

        8 & \emph{TP53, PIK3CA, MAP3K1, MAP2K4, PIK3R1, LPA, KRAS, ERBB2} & 0 & 0.004 & 0 \\ \hline

        9 & \emph{TP53, PIK3CA, MAP3K1, MAP2K4, PIK3R1, LPA, KRAS, ERBB2, FGFR2} & 0 & 0 & 0 \\ \hline

        10 & \emph{TP53, PIK3CA, MAP3K1, MAP2K4, PIK3R1, LPA, KRAS, ERBB2, FGFR2, TNXB} & 0 & 0 & 0 \\ \hline
      \end{tabular}
\begin{flushleft}
      $p_1$ and $p_2$ denote the $p$-values of the common gene sets in BRCA and OV, respectively. $p$ represents the overall significance.
\end{flushleft}
\end{table*}

We also employ BLP \cite{Zhao} to identify individual driver gene sets in BRCA and OV (Table \ref{table3}), respectively. For each $K$ in Table \ref{table3}, there is only one common gene \emph{TP53} between the identified gene sets for these two cancers. For other genes in the common gene sets in Table \ref{table2}, \emph{PIK3CA, MAP3K1, MAP2K4} and \emph{PIK3R1} only appear in the gene sets of BRCA, and \emph{KRAS, FGFR2} and \emph{LPA} appear only in those of OV (Table \ref{table3}). Thus, only a local path of the common mutated pathways (Fig. \ref{fig4}B) can be found by BLP for each of these two cancers (Fig. \ref{fig4}C).

\begin{table*}[!ht]
\renewcommand{\arraystretch}{1.3}
\caption{Significant individual driver gene sets in BRCA and OV}
\label{table3}
\footnotesize
\centering
      \begin{tabular}{|l|l|l|}
        \hline
        $K$ & Driver pathway in BRCA & Driver pathway in OV \\ \hline

        2 & \emph{TP53, PIK3CA} & \emph{TP53, KRAS} \\ \hline

        3 & \emph{TP53, PIK3CA, GATA3} & \emph{TP53, KRAS, IDI2} \\ \hline

        4 & \emph{TP53, PIK3CA, GATA3, CDH1} & \emph{TP53, KRAS, FGFR2, PIGV} \\ \hline

        5 & \emph{TP53, PIK3CA, GATA3, CDH1, CTCF} & \emph{TP53, KRAS, IDI2, PIGV, BRAF} \\ \hline

        6 & \emph{TP53, PIK3CA, GATA3, CDH1, CTCF, MACROD2} & \emph{TP53, KRAS, IDI2, BRAF, LPA, EGFR} \\ \hline

        7 & \emph{TP53, GATA3, CDH1, MACROD2, AKT1, MAP3K1, MAP2K4} & \emph{TP53, KRAS, IDI2, BRAF, PIGV, EGFR, C4orf45} \\ \hline

        8 & \emph{TP53, GATA3, CDH1, MACROD2, AKT1, MAP3K1, MAP2K4, PIK3R1} & \emph{TP53, KRAS, FGFR2, C4orf45, EPHA3, PPID, ETFDH, FNIP2} \\ \hline

        9 & \emph{TP53, GATA3, CDH1, MACROD2, AKT1, MAP3K1, MAP2K4, PIK3R1,} & \emph{TP53, KRAS, FGFR2, PIGV, EGFR, C4orf45, EPHA3, PPID,} \\
          & \emph{POLD4} & \emph{FNIP2} \\ \hline

        10 & \emph{TP53, GATA3, CDH1, MACROD2, AKT1, MAP3K1, MAP2K4, PIK3R1, } & \emph{TP53, KRAS, IDI2, BRAF, LPA, C4orf45, EPHA3,PPID, } \\
           & \emph{POLD4, ARID1A} & \emph{ETFDH, FNIP2} \\ \hline
      \end{tabular}
\begin{flushleft}
      Here the $p$-values are all less than 0.0001.
\end{flushleft}
\end{table*}

ComMDP has distinct advantages over both the gene-centric frequency-based approaches and the individual driver gene set based approaches. First, in the identified common gene sets (Table \ref{table2}), some genes have very low mutation frequency. For example, \emph{TNXB, LPA} and \emph{FGFR2} all have less than five mutations in 466 BRCA samples and 445 OV samples, respectively. With such low frequency, these genes cannot be discovered by the gene-centric frequency-based approaches. But all the three genes have important biological functions (Fig. \ref{fig4}B) and are closely related to the carcinogenesis of BRCA and OV \cite{Hu_2009,Kim_2010, Wang_2016,Campbell_2016}. For instance, Hu \emph{et al.} validated \emph{TNXB} as a promising biomarker for early metastasis of breast cancer \cite{Hu_2009}; Kim \emph{et al.} demonstrated \emph{TNXB} might be helpful to predict the prognosis of patients with stage III serous ovarian cancer through differential expression analysis \cite{Kim_2010}; \emph{LPA} and its receptors play an important role in mediating malignant behaviors in various cancers and recent studies \cite{ Wang_2016, Jesionowska} suggested they could be potential diagnostic biomarkers for BRCA and OV, respectively; \emph{FGFR2} were suggested as candidate targets for therapeutics in clinical trial for BRCA and OV \cite{Campbell_2016,Cole_2010}. Second, some of the ten important common genes (Table \ref{table2}) cannot be identified by the driver gene set identification approaches for individual cancer type (Table \ref{table3}). Especially, \emph{TNXB} and \emph{ERBB2} are not identified for any cancer by BLP. Actually, \emph{ERBB2} is a well-known cancer gene, and it plays a crucial role for certain subtypes of BRCA and OV patients \cite{Elizalde_2016,Hodeib_2015}. Third, it is important to note that the individual cancer type approach can only discover a small part of the common gene set for each cancer type (Fig. \ref{fig4}), whereas ComMDP can integrate information from different cancers and imply a more biologically reasonable common driver pathway.

\begin{figure*}[!t]
\centering
\includegraphics[width=5.5in]{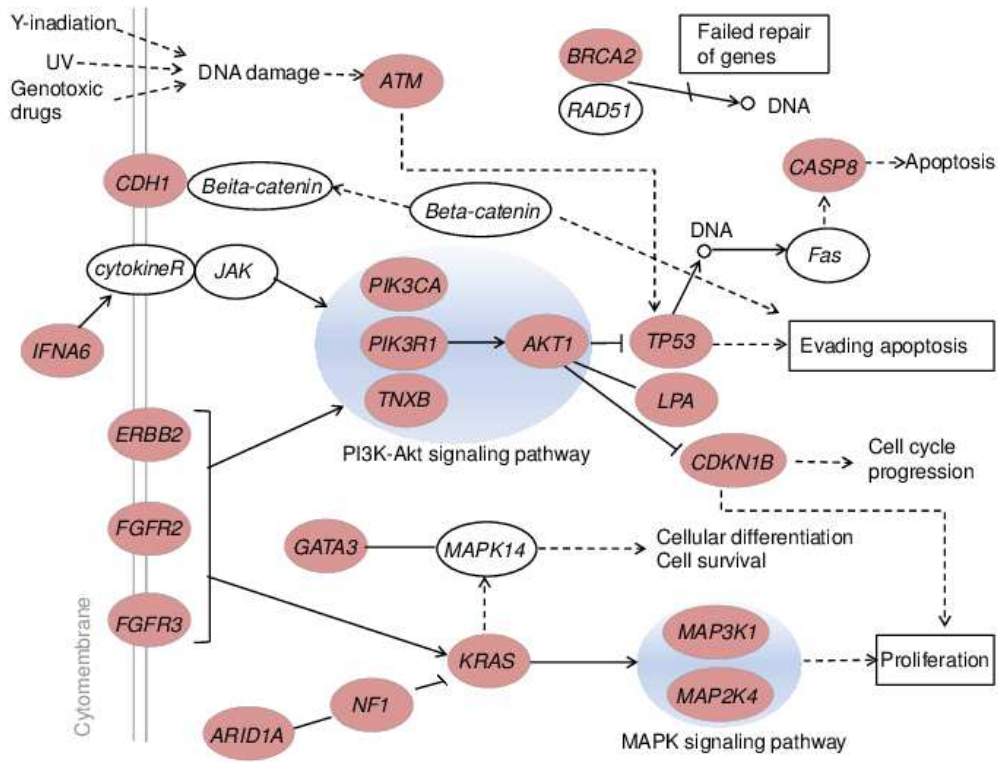}
\caption{Hypothetical driver pathways for BRCA carcinogenesis and progression. It is inferred based on the common driver gene sets between BRCA and other eight cancer types. The dotted arrows denote indirect effects, and a line represent a known interaction between them or co-occurrence in a known signaling pathway. The ComMDP discovered genes are in red.}
\label{fig5}
\end{figure*}

Importantly, identifying all the significant common driver gene sets of BRCA with certain cancer types will help to understand various aspects of BRCA carcinogenesis. Besides OV, other cancer types include BLCA, COADREAD, GBM, HNSC, KIRC, LUAD and UCEC (Suppl. Table S1). In total, we discover 38 different genes in all the eight significant gene sets (Suppl. Figure S2). These genes are involved in many important signaling pathways (Suppl. Figure S3) and relate to diverse cancers (such as prostate cancer, endometrial cancer, pancreatic cancer, lung cancer, glioma, colorectal cancer, etc) (searched by DAVID \cite{Huang_09a}). It is known that cancer is a very complex disease. We integrate prior pathway knowledge and all the common driver gene sets of BRCA with the eight cancer types to explore more details about BRCA carcinogenesis (Fig. \ref{fig5}). Compared to Fig. \ref{fig4}, we find some new paths and more genes involving in the important hallmarks of cancer in Fig. \ref{fig5}, such as [\emph{IFNA6--cytokineR/JAK}--PI3K-Akt signaling pathway--\emph{TP53--Fas--CASP8}] leading to apoptosis, [\emph{ARID1A--NF1--KRAS}--MAPK signaling pathway] leading to proliferation, [\emph{GATA3--MAPK14}] leading to cell survival, etc.

We note that, based on the common driver gene sets of BRCA with multiple cancer types, we can get distinct new discoveries versus previous work. For example, the authors in \cite{Kandoth} identified 127 significantly mutated genes (SMGs) from diverse signaling and enzymatic processes, and calculated the most frequently mutated genes in the pan-cancer cohort for each cancer type. Especially, for BRCA they obtained eight genes (\emph{TP53, PIK3CA, MAP3K1, MAP2K4, GATA3, AKT1, CDH1, CBFB}), seven of which belong to our 38 genes except \emph{CBFB}. On one hand, we find more genes involved in the cellular processes that the above eight genes relating to: transcription factor/regulator (\emph{CTCF}), genome integrity (\emph{ATM, BRCA2}), MAPK signaling (\emph{KRAS, NF1}), PI(3)K signaling (\emph{PTEN, PIK3R1}). On the other hand, we detect several genes involved in other important biological processes in cancer: histone modifier (\emph{ARID1A, PBRM1, KDM6A}), RTK signaling (\emph{FGFR2, FGFR3}), cell cycle (\emph{CDKN1B}). This indicates that these biological processes may also contribute to the carcinogenesis of BRCA. More importantly, we identify some other genes that are not included in the 127 selected genes in \cite{Kandoth}, but they play also crucial roles, such as \emph{CASP8, IFNA6, ERBB2, TNXB, NOTCH2} (Suppl. Figure S3). For example, \emph{CASP8} is involved in the programmed cell death induced by \emph{Fas} and various apoptotic stimuli, and there are many studies relating to its biological functions \cite{Kim_2016,Park_2016,Sagulenko}; \emph{NOTCH2} plays a role in a variety of developmental processes by controlling cell fate decisions, and has close relationship with BRCA progression \cite{KimRK_2016,Sehrawat}; \emph{IFNA6} belongs to the family of interferon, although it has not been well studied, this kind of immune-associated genes may be worth paying great attention for immunotherapy of cancers \cite{Niwakawa}.

Note that BLCA has common significant driver gene sets with all the other 10 cancer types (Suppl. Table S1). For BRCA stated above, some genes frequently appear in many common gene sets (Suppl. Figure S2). But for BLCA, the common gene sets are not necessarily the same, such as those with BRCA (Suppl. Table S2), COADREAD (Suppl. Table S3), GBM (Suppl. Table S4) and LUSC (Suppl. Table S5). For example, we identify two different sets of genes for BRCA and COADREAD, whereas their functional annotations are quite similar (Suppl. Figure S4). All these are closely related to cancer generation and progression.

\begin{figure}[!h]
	\centering
	\includegraphics[width=3.0in]{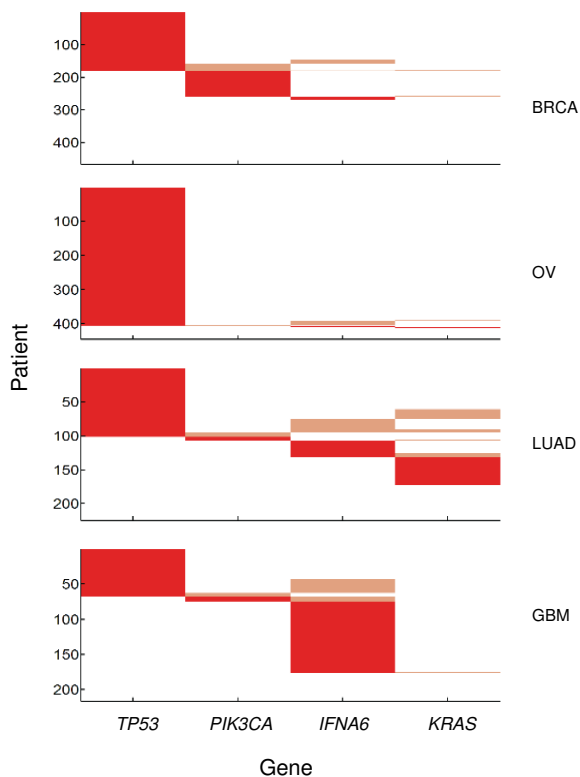}
	\caption{The heat map of the alterations in the common driver gene set (\emph{TP53, PIK3CA, IFNA6, KRAS}) among the four cancer types including BRCA, OV, LUAD and GBM.}
	\label{fig6}
\end{figure}

Furthermore, we also investigate the common mutated driver gene sets among multiple cancer types. For example, we find that BRCA, OV, LUAD and GBM have common significant gene sets with $K=$ 4 to 10 (Table \ref{table4}), which relate to the mutations of genes \emph{TP53, PIK3CA, KRAS, MAP3K1, EP300, PIK3R1, TNXB, KDM6A, LPA} and deletion of gene \emph{IFNA6}. As an example, we show the heat map of the alterations of the gene sets in these four cancer types for $K=4$ (Fig. \ref{fig6}). These gene alterations are approximately mutually exclusive in all the four cancer types. Compared to the situation of only considering BRCA and OV (Table \ref{table2}), it covers three new genes. Besides \emph{IFNA6} stated above, two others are \emph{EP300} and \emph{KDM6A}. \emph{EP300} interacting with \emph{TP53} is a transcriptional coactivator to mediate many transcriptional events including DNA repair \cite{Hasan}. It also functions as a histone acetyltransferase to regulate transcription via chromatin remodeling. Gene \emph{KDM6A} is associated with chromatin organization and transcriptional misregulation in cancer \cite{Su_2015}. Indeed, this investigation can help one to reveal common characteristics among diverse cancers.

\begin{table*}[!ht]
\renewcommand{\arraystretch}{1.3}
\caption{Significant common driver gene sets among BRCA, OV, LUAD and GBM}
\label{table4}
\small
\centering
      \begin{tabular}{|l|l|l|l|l|l|l|}
        \hline
        $K$ & Common pathway & $p_1$ & $p_2$ & $p_3$ & $p_4$ & $p$ \\ \hline

        4 & \emph{TP53, IFNA6, PIK3CA, KRAS} & 0.0010 & 0.0040 & 0.0020 & 0 & 0 \\ \hline

        5 & \emph{TP53, IFNA6, PIK3CA, KRAS, MAP3K1} & 0 & 0.0020 & 0.0040 & 0 & 0 \\ \hline

        6 & \emph{TP53, IFNA6, PIK3CA, KRAS, MAP3K1, EP300} & 0 & 0.0080 & 0.0040 & 0 & 0 \\ \hline

        7 & \emph{TP53, IFNA6, PIK3CA, KRAS, MAP3K1, EP300, PIK3R1} & 0 & 0.0040 & 0 & 0 & 0 \\ \hline

        8 & \emph{TP53, IFNA6, PIK3CA, KRAS, MAP3K1, EP300, PIK3R1, TNXB} & 0 & 0.0020 & 0.0010 & 0 & 0 \\ \hline

        9 & \emph{TP53, IFNA6, PIK3CA, KRAS, MAP3K1, EP300, PIK3R1, TNXB, KDM6A} & 0 & 0.0040 & 0.0020 & 0 & 0 \\ \hline

        10 & \emph{TP53, IFNA6, PIK3CA, KRAS, MAP3K1, EP300, PIK3R1, TNXB, KDM6A, LPA} & 0 & 0.0010 & 0.0080 & 0 & 0 \\ \hline
      \end{tabular}
\begin{flushleft}
      $p_1$, $p_2$, $p_3$ and $p_4$ denote the $p$-values of the common gene sets in BRCA, OV, LUAD and GBM, respectively. $p$ represents the overall significance.
\end{flushleft}
\end{table*}

\vskip5mm
\noindent{\it\textbf{Mutated driver gene sets specific to one cancer or multiple cancer types}}
\vskip2mm

We apply SpeMDP to the mutation data without common driver genes and identify several significant BRCA specific driver gene sets relative to OV with $K=$ 3, 4, 9, 10 (Table \ref{table5}). These gene sets relate to the mutations of \emph{GATA3, CDH1, AKT1, CTCF} and amplifications of \emph{ERBB2, WHSC1L1, CCND1, PLK1, RFPL4A, DDAH1}, many of which have been suggested to be closely related with breast cancer initiation and progression by a number of studies \cite{Huang_2015,Mustafa,Asp_2016,Irish}. For example, \emph{GATA3} plays a specific role in the differentiation of breast luminal epithelial cells, and has particular diagnostic utility in the setting of triple-negative breast carcinomas \cite{Asch-Kendrick}; the tumor suppressor \emph{CDH1} has been shown to be a potential drug target in breast cancer \cite{Huang_2015}; and epigenetic silencing of \emph{HOXA10} by \emph{CTCF} in breast cancer cells is related to tumorigenesis \cite{Mustafa}. Similarly, we also identify significant OV specific driver gene sets relative to BRCA with $K=$ 2 to 10 (Table \ref{table5}), and significant BRCA and OV specific driver pathways relative to the liquid cancer LAML with $K=$ 9, 10 (Suppl. Table S6).
\begin{table*}[!ht]
\renewcommand{\arraystretch}{1.3}
\renewcommand{\multirowsetup}{\centering}
\caption{BRCA and OV specific mutated driver gene sets relative to each other}
\label{table5}
\footnotesize
\centering
  	\begin{tabular}{|l|l|l|l|l|l|}
  		\hline
  Type	                & $K$ & Specific pathway & $p$ & $q$ & $P$ \\ \hline
  \multirow{4}{*}{BRCA/OV} & 3 & \emph{ERBB2, GATA3, CDH1} & 0.0110 & 1 & 0.0100 \\ \cline{2-6}
  		& 4 & \emph{ERBB2, GATA3, CDH1, WHSC1L1} & 0.0360 & 0.9510 & 0.0300 \\      \cline{2-6}
  		& 9 & \emph{ERBB2, GATA3, CDH1, WHSC1L1, CCND1, AKT1, CTCF, PLK1, RFPL4A} & 0.0410 & 0.7810 & 0.0420 \\  \cline{2-6}
  		& 10 & \emph{ERBB2, GATA3, CDH1, WHSC1L1, CCND1, AKT1, CTCF, PLK1, RFPL4A, DDAH1} & 0.0160 & 0.7550 & 0.0280 \\ \cline{2-6}
      	\hline
  \multirow{9}{*}{OV/BRCA} & 2 & \emph{BRCA1, BRCA2} & 0 & 0.6000 & 0 \\ \cline{2-6}
  		& 3 & \emph{BRCA1, BRCA2, CACNA1A} & 0 & 0.6570 & 0 \\   \cline{2-6}
  		& 4 & \emph{BRCA1, BRCA2, CACNA1A, WT1} & 1.0000e-03 & 0.6770 & 0 \\  \cline{2-6}
  		& 5 & \emph{BRCA1, BRCA2, CACNA1A, CASC1, GUSBP3} & 0 & 1 & 0 \\  \cline{2-6}
  		& 6 & \emph{BRCA1, BRCA2, CACNA1A, CASC1, GUSBP3, HUS1B} & 0 & 1 & 0 \\  \cline{2-6}
  		& 7 & \emph{BRCA1, BRCA2, WT1, ADPRHL2, METTL17, DNM2, COX4I2} & 0 & 0.9970 & 0 \\ \cline{2-6}
  		& 8 & \emph{BRCA1, BRCA2, WT1, ADPRHL2, METTL17, DNM2, COX4I2, SRP19} & 0 & 1 & 0 \\ \cline{2-6}
  		& 9 & \emph{BRCA1, BRCA2, WT1, ADPRHL2, METTL17, DNM2, COX4I2, SRP19, PARP8} & 0 & 1 & 0 \\ \cline{2-6}
  		& 10 & \emph{BRCA1, BRCA2, WT1, ADPRHL2, METTL17, DNM2, COX4I2, SRP19, PARP8, PRPS2} & 0 & 1 & 0 \\ \hline
  	\end{tabular}
\begin{flushleft}
  	  $p$ and $q$ denote the $p$-values of the gene set in BRCA relative to OV (BRCA/OV) or vice versa (OV/BRCA), respectively. $P$ represents the overall significance. Here the identified gene set is significant means that $p$ and $P$ are both less than 0.05, but $q$ is larger than 0.05.
\end{flushleft}
\end{table*}

\vskip5mm
\noindent{\it\textbf{The liquid cancer LAML has significant common or specific driver gene sets compared to solid cancer types}}
\vskip2mm

LAML is the only liquid cancer in the current study. Interestingly, it has some common driver gene sets with solid cancers. Specifically, by using ComMDP we identify LAML has a significant common driver gene set with COADREAD, GBM and BLCA for $K = 5$, which includes deletion of \emph{IFNA6}, and mutations of \emph{TP53, IDH1, WT1, SDK1}.

\begin{figure}[!t]
	\centering
	\includegraphics[width=3.0in]{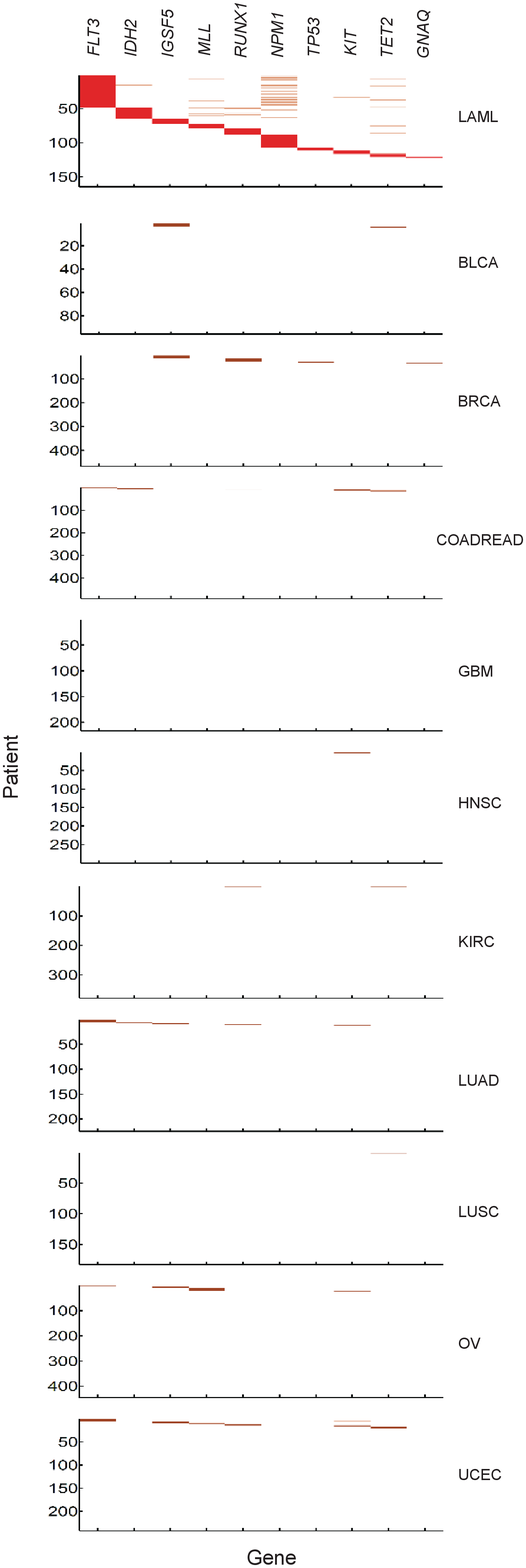}
	\caption{The heat map of the alterations in the LAML specific driver gene set (\emph{FLT3, IDH2, IGSF5, MLL, RUNX1, NPM1, TP53, KIT, TET2, GNAQ}) relative to other ten cancer types including BLCA, BRCA, COADREAD, GBM, HNSC, KIRC, LUAD, LUSC, OV and UCEC.}
	\label{fig7}
\end{figure}

\begin{figure*}[!t]
\centering
\includegraphics[width=5.2in]{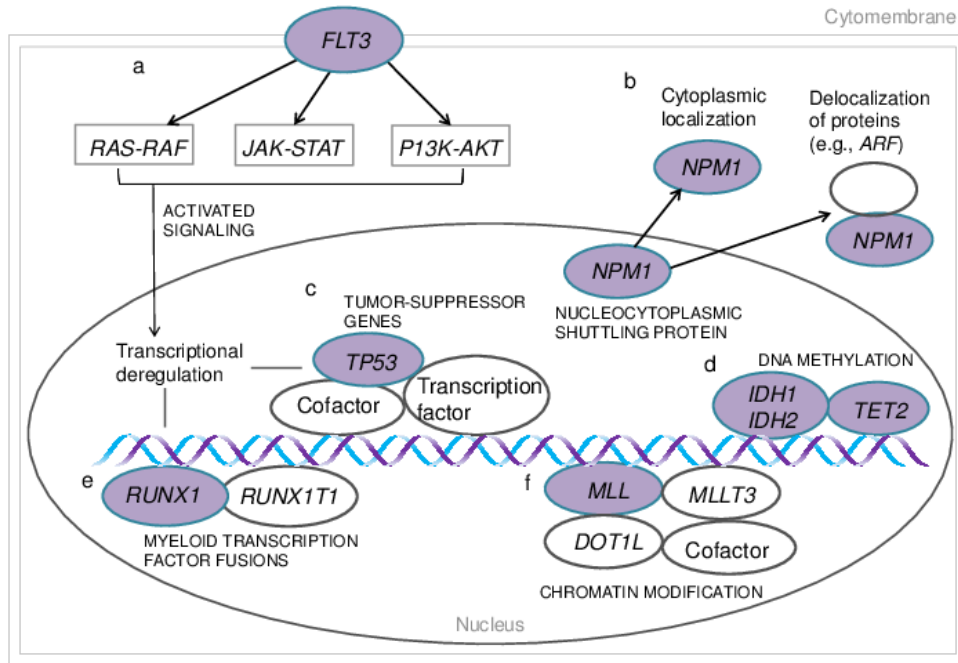}
\caption{Specific driver pathways or biological processes of LAML relative to the 10 solid cancer types. (a) Mutations in \emph{FLT3} confer a proliferative advantage through the \emph{RAS-RAF}, \emph{JAK-STAT}, and \emph{PI3K-AKT} signaling pathways. (b) Mutations in \emph{NPM1} result in the aberrant cytoplasmic localization of \emph{NPM1} and \emph{NPM1}-interacting proteins. (c) Deletions of tumor suppressor genes, such as \emph{TP53}, lead to transcriptional deregulation and impaired degradation through \emph{MDM2} and \emph{PTEN}. (d) \emph{DNMT3A} and \emph{TET2} mutations, as well as \emph{IDH1} and \emph{IDH2} mutations, can lead to the deregulation of DNA methylation. (e) Mutations in myeloid transcription factors such as \emph{RUNX1} and transcription factor fusions by chromosomal rearrangements lead to transcriptional deregulation and impaired hematopoietic differentiation. (f) Mutations of genes involved in the epigenetic homeostasis of cells, such as mutations of \emph{ASXL1} and \emph{EZH2}, lead to deregulation of chromatin modification as well as \emph{MLL-MLLT3} gene fusion, which can impair other methyltransferases. Note: the genes in purple represent they appear in the identified specific driver gene sets. (Referring to \cite{Dohner,TCGA_2013})}
\label{fig8}
\end{figure*}

More importantly, LAML is expected to have some specific mutation patterns. We investigate LAML specific driver pathways relative to other 10 solid cancers, and discover significant driver gene sets with $K = 2$ to 10 except $K = 5$ (Table \ref{table6}). These gene alterations include mutations of \emph{FLT3, IDH2, NRAS, IDH1, RUNX1, NPM1, TET2, KIT}, amplifications of \emph{MLL, IGSF5} and deletions of \emph{TP53, GNAQ}. We show the heat map of the alterations of the gene sets in all the 11 cancer types for $K=10$ (Fig. \ref{fig7}) and see that the alterations display significant mutual exclusivity in LAML, but not in other ten cancer types. Most of these identified genes have been previously reported to be related to LAML \cite{Dohner}. For example, eight genes of them are involved with six functional categories associated with LAML carcinogenesis (Fig. \ref{fig8}). Many large-scale studies have confirmed that \emph{FLT3} can activate mutations in LAML occurrence and disease progression and thus plays an important role in the pathogenesis of LAML \cite{Dohner, Gale}; \emph{NPM1} is thought to be involved in several processes including centrosome duplication, cell proliferation and regulation of the \emph{ARF/TP53} pathway and its mutations are associated with LAML supported by various studies \cite{ Hefazi,Alpermann,Chiu}; \emph{KIT} confers unfavorable prognosis for LAML patients \cite{Dohner}.

\begin{table*}[!ht]
\renewcommand{\arraystretch}{1.3}
\caption{LAML specific mutated driver gene sets relative to BRCA, HNSC, KIRC, LUSC, BLCA, GBM, LUAD, COADREAD, OV and UCEC}
\label{table6}
\small
\centering
      \begin{tabular}{|l|l|l|l|l|}
        \hline
        $K$ & Specific pathway & $p$ & $q_1, \cdots, q_{10}$ & $P$ \\
        \hline
        2 & \emph{FLT3, IDH2} & 0.0150 & 1.0000, 1.0000, 1.0000, 1.0000, 1.0000, & 0.0170 \\
          &  &  & 1.0000, 0.9520, 0.9890, 1.0000, 1.0000 &  \\
        \hline
        3 & \emph{FLT3, IDH2, NRAS} & $<0.0001$ & 1.0000, 1.0000, 1.0000, 1.0000, 1.0000, & 0.0010 \\
          &  &  & 1.0000, 0.9220, 0.7220, 0.9960, 0.8160 &  \\
        \hline
        4 & \emph{FLT3, IDH2,} & 0.0020 & 1.0000, 1.0000, 1.0000, 1.0000, 0.9480, & 0.0020 \\
          & \emph{NRAS, IDH1} &  & 0.9340, 0.8970, 0.5450, 0.9960, 0.8450 &  \\
        \hline
        6 & \emph{FLT3, MLL, IGSF5,} & 0.0180 & 0.5110, 1.0000, 1.0000, 1.0000, 1.0000, & 0.0140 \\
          & \emph{RUNX1, NPM1, TP53} &  & 1.0000, 0.9920, 0.9830, 0.9840, 0.9640 &  \\
        \hline
        7 & \emph{FLT3, IDH2, IGSF5,} & 0.0430 & 0.4720, 0.9940, 0.9930, 1.0000, 0.9630, & 0.0240 \\
          & \emph{RUNX1, NPM1, TP53, TET2} &  & 1.0000, 0.9930, 0.9930, 0.9800, 0.9410 &  \\
        \hline
        8 & \emph{FLT3, IDH2, IGSF5, MLL,} & 0.0050 & 0.5020, 0.9950, 1.0000, 1.0000, 1.0000, & 0.0050 \\
          & \emph{RUNX1, NPM1, TP53, KIT} &  & 1.0000, 0.9630, 0.8860, 0.9640, 0.9870 &  \\
        \hline
        9 & \emph{FLT3, IDH2, IGSF5, MLL,} & 0.0130 & 0.4890, 0.9900, 0.9960, 1.0000, 0.9710, & 0.0050 \\
          & \emph{RUNX1, NPM1, TP53, KIT, TET2} &  & 1.0000, 0.9650, 1.0000, 0.9680, 0.9950 &  \\
        \hline
        10 & \emph{FLT3, IDH2, IGSF5, MLL, GNAQ,} & 0.0090 & 0.3820, 0.9780, 0.9980, 1.0000, 0.9750, & 0.0030 \\
           & \emph{RUNX1, NPM1, TP53, KIT, TET2}  &  & 1.0000, 0.9790, 1.0000, 0.9740, 0.9880 &  \\
        \hline
      \end{tabular}
\begin{flushleft}
$p$, $q_1, \cdots, q_{10}$ denote the $p$-values of the gene set in LAML, BRCA, HNSC, KIRC, LUSC, BLCA, GBM, LUAD, COADREAD, OV and UCEC, respectively. $P$ represents the overall significance. For each $K$, $p$ and $P$ are less than 0.05, but $q_1, \cdots, q_{10}$ are all larger than 0.05.
\end{flushleft}
\end{table*}

Moreover, we can predict the potential implication of \emph{GNAQ} with LAML based on its appearance in the LAML specific driver gene set even with very low mutation frequency in LAML (2/164). \emph{GNAQ} has been considered as one of uveal melanoma driver genes \cite{Royer-Bertrand}, and a prognostic factor for mucosal melanoma \cite{Sheng_2016}. Another study \cite{Marjanovic} indicates that variations of \emph{GNAQ} tend to occur in childhood LAML patients. On the other hand, mutational-driven comparison with other cancer types showed that uveal melanoma is very similar to pediatric cancers, characterized by very few somatic insults and, possibly, important epigenetic changes \cite{Royer-Bertrand}. Thus, we suggest that \emph{GNAQ} might be a candidate driver gene for childhood LAML patients and its function in LAML carcinogenesis and progression worth further exploration.

\section{DISCUSSION}

In this study, we develop ComMDP and SpeMDP to identify cancer common and specific mutated driver gene sets among two or multiple cancer types, respectively. We first apply them to a set of simulated data with diverse mutation rates and pathway sizes to demonstrate their effectiveness. We further apply them to real biological data from TCGA, and obtain a set of cancer common and specific gene sets which are involved in several key biological processes or signaling pathways. This suggests that the identified common or specific driver gene sets may play crucial roles and worth to be further explored.

Applications of ComMDP and SpeMDP to real data show their advantages over both gene-centric frequency-based approaches and individual driver gene set based approaches. For example, we identified \emph{TNXB, LPA, FGFR2, CASP8, NOTCH2} for BRCA, all of which are mutated with very low frequency (less than five mutations in 466 patients), but have critical biological functions in carcinogenesis of BRCA. All these genes cannot be discovered by the gene-centric frequency-based approaches \cite{Kandoth}. We also find that some of the identified important common genes (Table \ref{table2}) cannot be detected by the driver gene set identification approaches for individual cancer types \cite{Zhao}. Moreover, the individual cancer type approaches can only discover a small part of the common pathway for each cancer type (Fig. \ref{fig4}), whereas ComMDP can integrate information from different cancers and give a more biologically reasonable common driver pathway. Furthermore, in the specific driver gene sets of LAML relative to solid cancer types by SpeMDP, \emph{GNAQ} (with only two mutations in 164 LAML patients) has showed potential implication with LAML carcinogenesis and progression, but it cannot be detected by gene-centric approaches.

We obtain the common driver gene sets of all pairs of the 11 cancer types with $K=$ 2 to 10 (Suppl. Table S1), and note that the significance of common driver gene sets has no transitivity. For example, there are significant common driver gene sets between LAML and COADREAD as well as LAML and LUAD for $K=$ 3 to 10 (Suppl. Table S7 and S8). But there are no significant ones between COADREAD and LUAD for $K=$ 2 to 10. In contrast, there are no significant common gene sets between GBM and HNSC as well as HNSC and OV for $K=$ 2 to 10, but there are significant ones between GBM and OV for $K=$ 3 to 10 (Suppl. Table S9).

In this study, to identify common driver gene sets, we first select the genes which have mutations in all the examined cancer types for further analysis. In fact, this model can be generalized to include the genes which have no mutations in some of the considered cancer types. We may add some constrains to ensure that the number of non-mutation cancer types is not more than a preassigned number for any considered gene. Moreover, it can also be used to investigate the commonalities and specificities among different subtypes within a certain cancer. We expect that our methods can provide crucial information for understanding the molecular mechanism of cancer generation and progression.

\section{AVAILABILITY}

The methods are implemented in the MATLAB code and are available upon request.

\section{SUPPLEMENTARY DATA}

Supplementary Data are available at NAR Online.

\section{ACKNOWLEDGEMENTS}

We would like to thank Dr. Yong Wang from the Academy of Mathematics and Systems Science, CAS for his constructive comment.

\section{FUNDING}

This work was supported by the National Natural Science Foundation of China [No. 61379092, 61422309, 61621003 and 11661141019]; the Strategic Priority Research Program of the Chinese Academy of Sciences (CAS) [XDB13040600], the Outstanding Young Scientist Program of CAS, CAS Frontier Science Research Key Project for Top Young Scientist [No. QYZDB-SSW-SYS008], and the Key Laboratory of Random Complex Structures and Data Science, CAS [No. 2008DP173182].


\end{document}